\documentclass{article}
\usepackage{authblk}
\usepackage{lipsum}
\usepackage{setspace}
\usepackage{indentfirst}
\usepackage{xcolor}
\usepackage{pifont}
\usepackage{rotating}
\usepackage{lscape}
\usepackage[lmargin=2cm,rmargin=2cm,tmargin=2cm,bmargin=2cm]{geometry}
\usepackage[labelformat=simple]{subcaption}
\usepackage{float}
\usepackage{booktabs}
\usepackage{lscape}
\usepackage{rotating}
\usepackage{caption}
\usepackage{subcaption}
\usepackage{longtable}
\usepackage{multicol}
\usepackage{sectsty}
\usepackage{titlesec}
\usepackage{natbib}
\usepackage{graphicx}
\usepackage{float}
\usepackage{epstopdf}
\usepackage{tikz}
\usepackage{xcolor}
\usepackage{amsmath}
\usepackage[labelformat=simple]{subcaption}
\usepackage{hyperref}
\usepackage[T1]{fontenc}
\sectionfont{\fontsize{11}{13}\selectfont}
\subsectionfont{\fontsize{10}{12}\selectfont}
\subsubsectionfont{\fontsize{9}{11}\selectfont}
\titlelabel{\thetitle.\quad}
\title{\textbf{Multifractal behavior of price changes in the
Green Bonds funds}} 
\author[1]{WENDERSON GOMES BARBOSA}
\author[2]{KEROLLY KEDMA FELIX DO NASCIMENTO}
\author[3]{FÁBIO SANDRO DOS SANTOS}
\author[4]{SILVIO FERNANDO ALVES XAVIER JUNIOR}
\author[1]{TIAGO A. E. FERREIRA}

\affil[1]{Department of Statistics and Informatics, Federal Rural University of Pernambuco, Dom Manuel de Medeiros, N/A \\
Recife-PE, 5217-900, Brazil\\
wenderson.gomes@ufrpe.br}

\affil[2]{Department of mathematics, Regional University of Carriri, Leão Sampaio Avenue,107\\
Juazeiro do Norte-CE, 63041-235,Brazil\\
kerollyfn@gmail.com}

\affil[3]{Department of Agronomy, Federal University of Piauí, BR-135, KM 3, \\
Bom Jesus-PI, 64900-000,
Brazil \\
fabio.santos@ufpi.edu.br}

\affil[4]{Department of Statistics, State University of Paraiba, Baraúnas Street, 351\\
Campina Grande-PB, 58429-500,
Brazil\\
silvio@servidor.uepb.edu.br}

\affil[5]{Department of Statistics and Informatics, Federal Rural University of Pernambuco, Dom Manuel de Medeiros, N/A \\
Recife-PE, 5217-900, Brazil\\
tiago.espinola@ufrpe.br}

\begin{document}
\maketitle
\begin{abstract}
Climate change has driven the market to seek new ways of raising funds to mitigate its effects. One such innovation is the emergence of Green Bonds — financial assets specifically designed to support sustainable projects. This study explores the fractal behavior of daily price changes in thirty-five Green Bond funds using the Multifractal Detrended Fluctuation Analysis (MFDFA) method. Our results indicate that price changes exhibit persistent behavior and high multifractality, characterized by large fluctuations. Only one of the thirty-five time series analyzed showed an outlier result, suggesting that the funds display very similar behavior. By shuffling the series, we were able to reduce multifractality significantly. These findings suggest that Green Bond funds exhibit multifractal behavior typical of other financial assets.
\end{abstract}
\section{INTRODUCTION}\label{sec:intro}
Green bonds are being studied primarily using econometric methods (Jian, 2023)\cite{jian2023green}. Jian (2023) employed self-fulfilling vectors in a panel data analysis of Green Bonds from several countries that adopt them, confirming the relevance of green bonds in the European Commission’s investments.  Using the Quantile-on-Quantile method used by Kartal \textit{et al.}\cite{Kartal2025} The Green Bonds are effective in reducing $CO_{2}$  emissions in China and the USA, and in China, they can contribute to the growth of solar and wind energy. 
The world has been undergoing significant transformations since the discovery that the planet's climate is changing \cite{calvet2022finance} \cite{azizi2023noninteger}. Since then, alternatives have been sought to mitigate this fact. For this, the European Investment Bank created the first Green Bonds to raise funds for investments that do not seek to cause or reduce environmental damage~\cite{ebeling2022european}. It is still poorly understood how these green securities react to market dynamics, as they are a new type of asset, similar to cryptocurrencies. As a result of the worsening climate change, Green Bonds and their derivatives, such as funds and ETFs (Exchange-Traded Funds), have gained increasing attention in the financial market~\cite{Michetti2023}. Despite the growing demand for such work, there is a lack of a broad understanding of empirical aspects, such as the behavior of complex systems, which creates a gap in the literature. 

Green Bonds are primarily being studied using econometric methods 
\cite{jian2023green}, where it was employed self-fulfilling vectors in a panel data analysis of Green Bonds from several countries that adopt them, confirming the applicability of Green Bonds in the European Commission's investments. 
Using the Quantile-on-Quantile method, Kartal et al.~\cite{kartal2024dynamic} found that Green Bonds effectively reduce $CO_2$ emissions in China and the US, and in China, they can contribute to the growth of solar and wind power. 
Analyzed with the ADF and BDS tests and the WLMC model, in the Green Bonds index of the S\&P Green Bond Index, the Price of Crude Oil (COP), the Global Geopolitical Risk index (GPR) and managed to demonstrate that Green Bonds can impact $CO_{2}$ emissions. The studies demonstrate the importance of performing statistical analysis and show that its results can have a positive impact on society. 

Several methods are used to understand the complex behavior of non-stationary time series. The fractal and multifractal approach is widely used to investigate complex phenomena. The MFDFA tool~\cite{kantelhardt2002multifractal} is one of the most powerful for this type of analysis.
The literature contains works that address other financial assets from the fractal and multifractal perspective \cite{liu1999statistical,zunino2008multifractal,jiang2011multifractal,stovsic2015multifractal2,li2020institutional,nejad2021multifractal,moudud2024stock}. Within specifically in the category of financial assets studies by this method, we can consider cryptocurrencies~\cite{lahmiri2018chaos,stosic2019multifractal}, commodities~\cite{stosic2020multifractal,nascimento2022covid}, and market sectors~\cite{stosic2019multifractalmarket} which has been gaining increasing importance in the literature due to its significant demand, having a non-persistent multifractal behavior of small fluctuations, very close to the stock market.

In this article, we studied the multifractal parameters of the logarithmic returns of the prices of thirty-five Green Bond Funds. The MFDFA method was applied to examine the correlations of price changes. To understand the behavior of time series, it is essential to recognize that they exhibit a pattern over time.
The article is organized as follows. Section \ref{sec:methods} presents the MFDFA method used in this article, along with the data set. Section \ref{sec:result} describes the results obtained. Section \ref{sec:concluse} presents the conclusion of the analysis.

\section{METHODS}\label{sec:methods}

\subsection{Multifractal Detrended Fluctuation Analysis}

Diverse methods have been proposed to analyze the multifractal behavior of time series \cite{de2025multifractal} \cite{liu2025analysis} \cite{rahmani2025unveiling}.  The Multifractal Detrended Fluctuation Analysis (MFDFA)~\cite{kantelhardt2002multifractal} method is a generalization of the Detrended Fluctuation Analysis (DFA) method and has been demonstrated to be a robust instrument to analyze the multifractal behavior. In this article,
we apply the MFDFA method, which has been successfully applied in various areas of knowledge, such as in financial series applications~\cite{matia2003multifractal, zunino2008multifractal,oh2012multifractal,stovsic2015multifractal,stosic2019multifractal,choi2021analysis}, energy market~\cite{wang2022impact}, physiological signals~\cite{figliola2007multifractal}, hydrological processes~\cite{kantelhardt2006long}, geophysical~\cite{telesca2006measuring}, and solar radiation \cite{akinsusi2022nonlinear,dos2023multifractal}. For the implementation of the method, we determine the following stages~\cite{kantelhardt2002multifractal} of the MFDFA:

\begin{itemize}
\item Obtain the accumulated integral of the series $x(i)$, $i = 1,..., N$ to produce $X(k)$ = $\sum^{k}_{i=1}[x(i)-\bar{x}]$, where $\bar{x}$ is the average.
\item Perform series segmentation $X(k)$ into $N_n$ non-overlapping segments of equal lenfth $n$ and estimate the local trend $X_i(k)$ in each segment from a mth order polynomial regression.
\item  Remove the local trend from the integrated series in each segment (by subtracting the local trend) to determine the order fluctuation function $q$ by calculating the average variance in all segments.
$$
F_{q}(n) = \Bigg\{\frac{1}{N_{n}}\sum_{i=1}^{N_{n}}\bigg[\frac{1}{2}\sum^{in}_{k=(i-1) n+1}[X(k) - X_{i}(k)]^{2}\bigg]^{\frac{q}{2}}\Bigg\}^{\frac{1}{q}}
$$

where $q$ can take any real value except zero.
\item  Repeat the procedure for all box sizes $n$ to obtain the floating function $F_{q}(n)$. The presence of long-range correlations is verified if $F_{q}(n)$  will increase with $n$ as a power law $F_{q}(n) \sim n^{h(q)} $, At this point, the scale exponent $h(q)$ is determined by the slope of the linear regression of log $F_{q}(n)$ as a function of $ log n$. As the scale exponent $h(2)$ is equivalent to the well-known Hurst exponent for stationary time series, $h(q)$ is designated as the generalized Hurst exponent.
\end{itemize}

In multifractal processes, the generalized Hurst exponent is a decreasing function of $q$ and describes the scaling behavior of large (small) fluctuations for positive (negative) $q$ values. The exponent relates to the classical multifractal exponent defined by the standard partition function multifractal formalism as $\tau(q) = qh(q)$, where $\tau(q)$ is a linear function for monofractal signals and a nonlinear function for multifractal signals.The singularity exponent $\alpha$ characterizes local scaling behavior, while the multifractal spectrum $f(\alpha)$is derived via the Legendre transform.

$$
\alpha(q) = d\tau(q)/dq, ~f(\alpha) = q\alpha - \tau(q),
$$
where $f(\alpha)$ denotes the fractal dimension of the series subset characterized by the H\"older exponent $\alpha$. For monofractal signals, the singularity spectrum produces a single point in the $f(\alpha)$ plane, whereas multifractal processes yield a bell-shaped spectrum. 

We fit the singularity spectra to a fourth-degree polynomial to measure the complexity of the series with

$$
f(\alpha) = A + B(\alpha -\alpha_0) + C(\alpha -\alpha_0)^2 + D(\alpha -\alpha_0)^3 + E(\alpha - \alpha_0)^4
$$
and we calculate a set of multifractal spectrum parameters: the maximum position $\alpha(0)$; the width of the spectrum $W$ (doing $\alpha_{max} - \alpha_{min}$).
Obtained from extrapolating the fitted curve to zero and the skew parameter
$$
r = \frac{\alpha_{max} - \alpha_{0}}{\alpha_{0} -\alpha_{min}} ~ ,
$$
where $r = 1$ for symmetric shapes, $r > 1$ right- skewed shapes, and $r < 1$ for left-skewed shapes. Generally speaking, a small value of $\alpha_0$ suggests that the series is correlated and more regular in appearance. The width of the spectrum $W$ measures the degree of multifractality in the series (the wider the range of fractal exponents, the richer the structure of the series). The skew parameter $r$ determines which fractal exponents are dominant: fractal exponents that describe the scaling of small fluctuations for the right-skewed spectrum and fractal exponents that describe the scaling of large fluctuations for the left-skewed spectrum. These three parameters$(\alpha_0, W ,r)$ lead to a measure of complexity in which a series with a high value of $\alpha_0$, a wide range $w$ of scaling exponents, and a right-skewed shape can be considered more complex than one with opposite characteristics. 

\subsection{Data}
\begin{figure}[H]
    \centering
    \begin{subfigure}{0.49\textwidth}
        \centering
        \includegraphics[width=\linewidth]{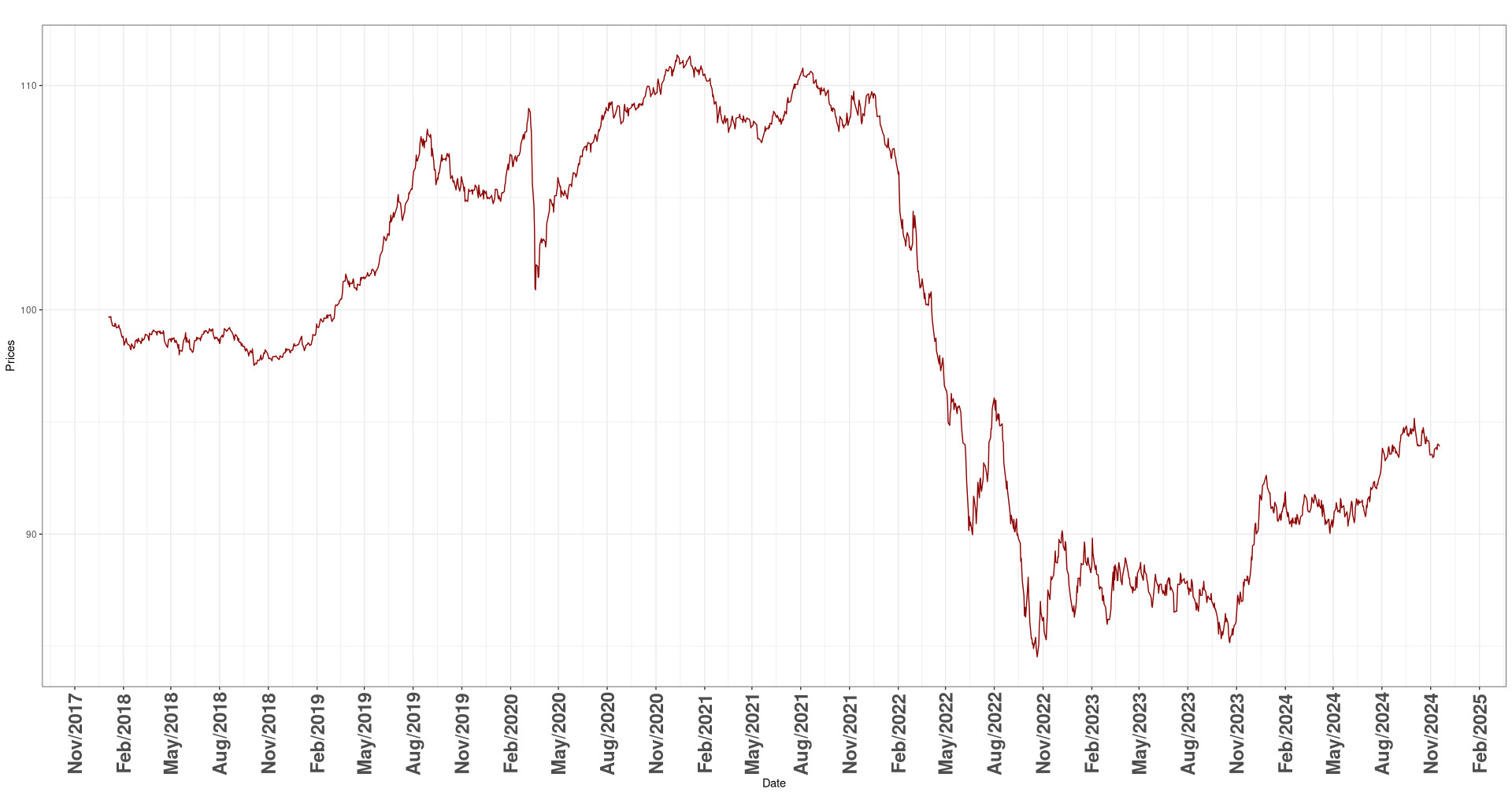}
        \caption{Amundi Responsible Investing - Impact Green Bonds DP }
        \label{fig:fig1}
    \end{subfigure}
    \hfill
    \begin{subfigure}{0.49\textwidth}
        \centering
        \includegraphics[width=\linewidth]{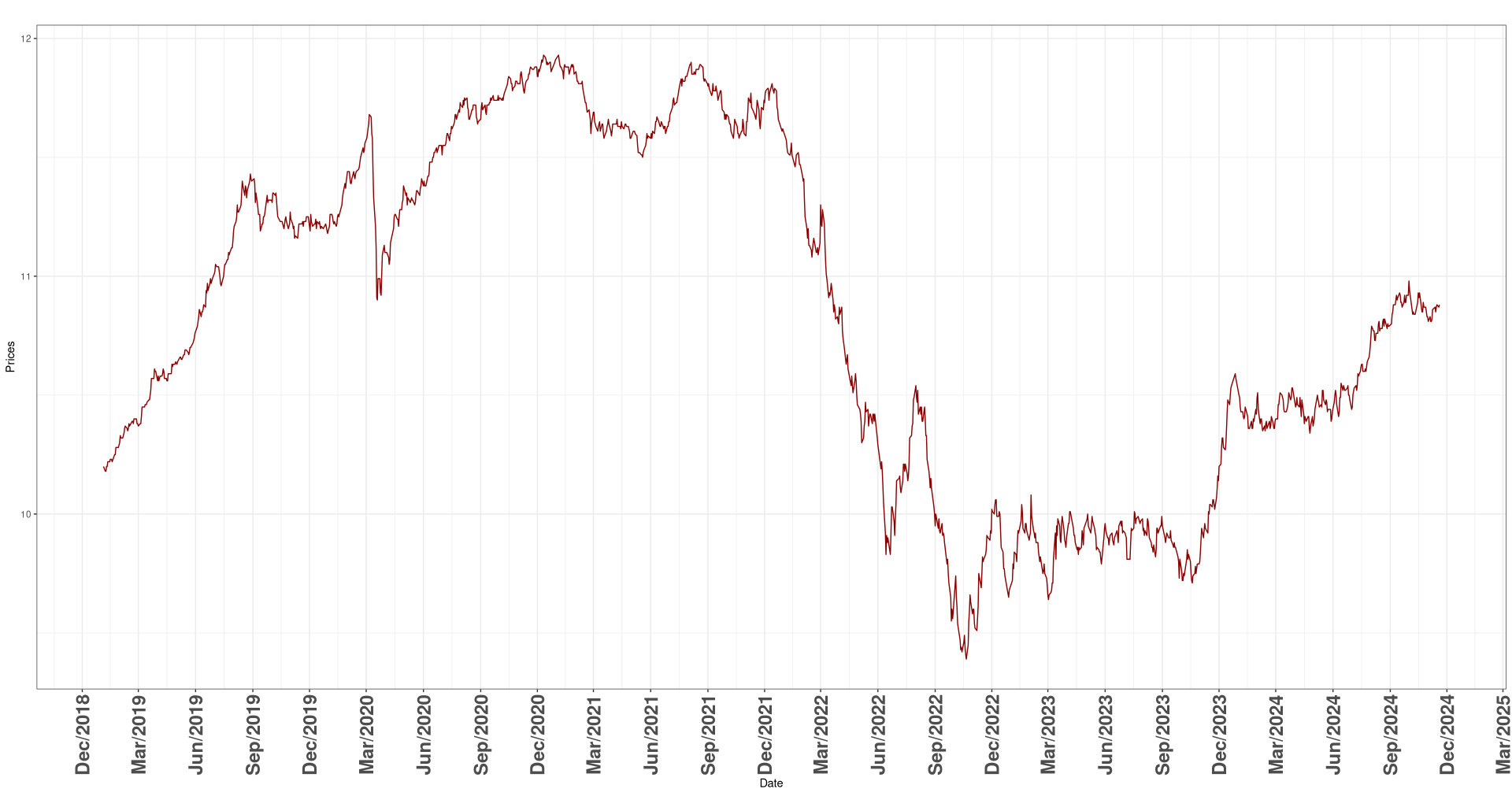}
        \caption{iShares Green Bond Index Fund (IE) D Acc USD Hedged}
        \label{fig:fig2}
    \end{subfigure}
    
    \begin{subfigure}{0.49\textwidth}
        \centering
        \includegraphics[width=\linewidth]{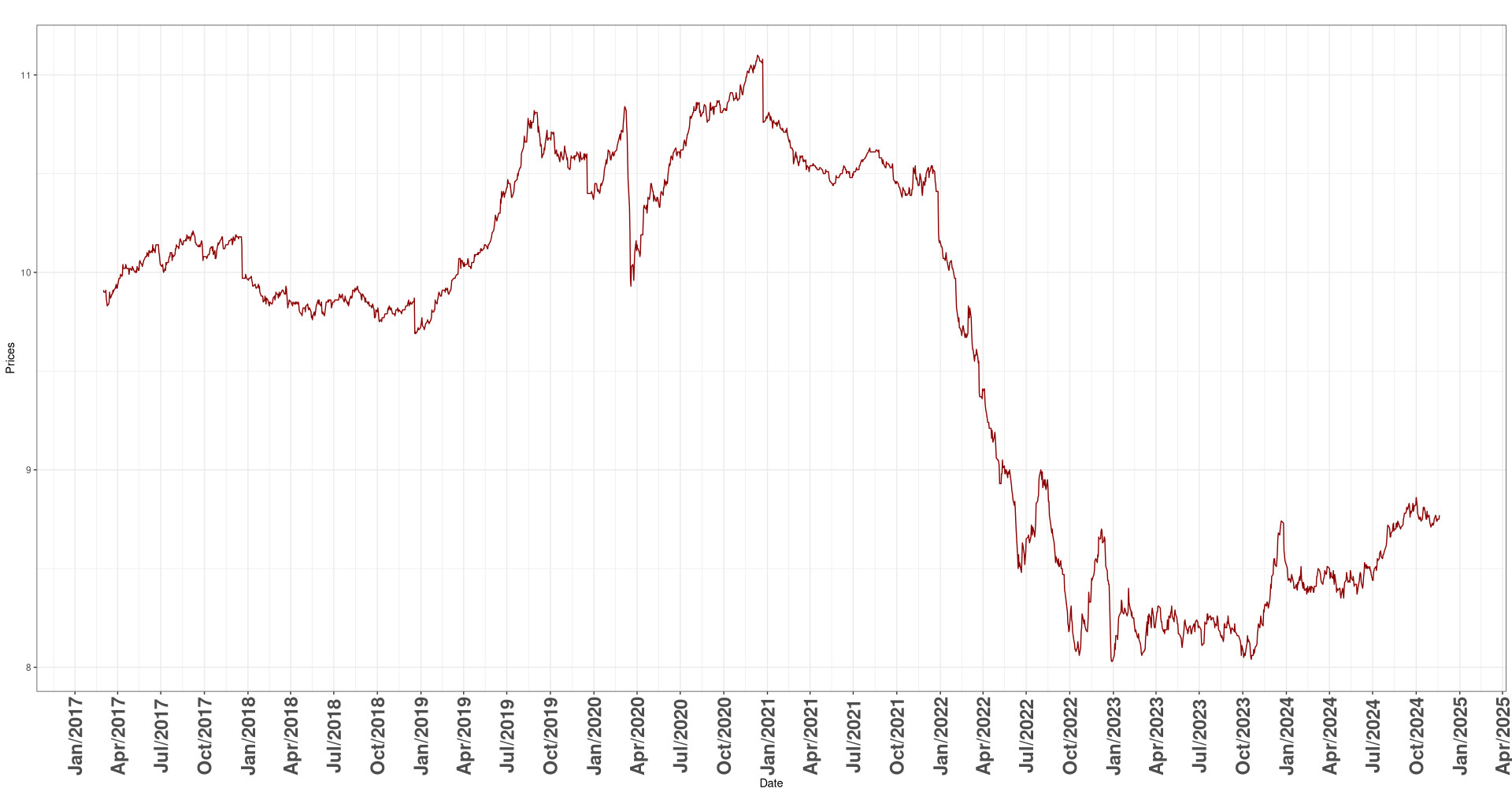}
        \caption{Mirova Global Green Bond Fund Class Y }
        \label{fig:fig3}
    \end{subfigure}
    \hfill
    \begin{subfigure}{0.49\textwidth}
        \centering
        \includegraphics[width=\linewidth]{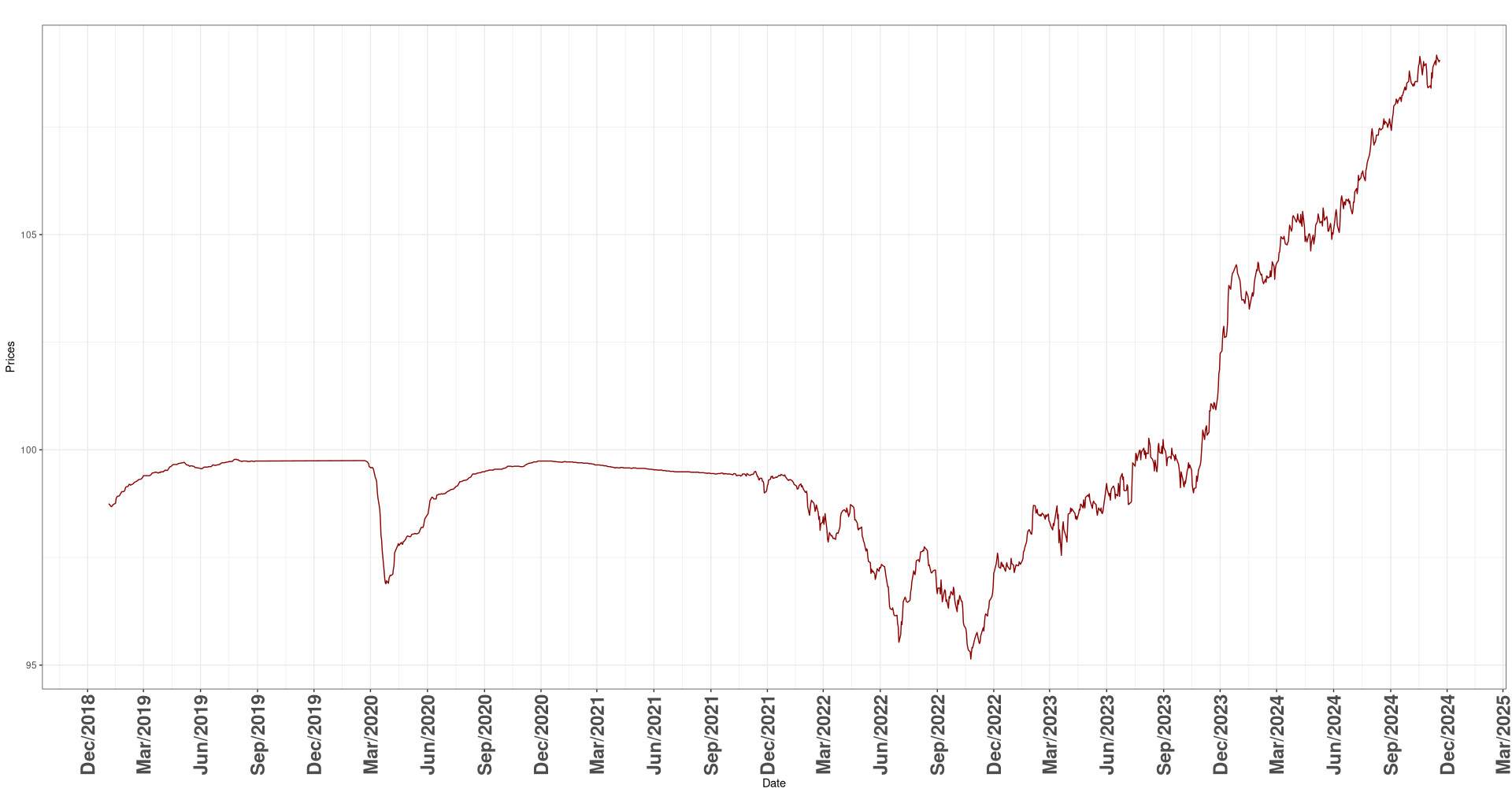}
        \caption{Swiss Life Funds (F) Bond Floating Rates F}
        \label{fig:fig4}
    \end{subfigure}
    
    \caption{Graphs of the four Green Bond Funds prices for the observation series.}
    \label{fig:st}
\end{figure}

In this article, we investigate the daily closing prices 35 Green Bonds funds listed in Table \ref{tab:tab1}. Data were obtained from \href{https://finance.yahoo.com/}{Yahoo Finance} from November 04, 2013 to November 19, 2024. Figure \ref{fig:st} shows four Green Bond Funds.

\begin{figure}[H]
    \centering
    \begin{subfigure}{0.49\textwidth}
        \centering
        \includegraphics[width=\linewidth]{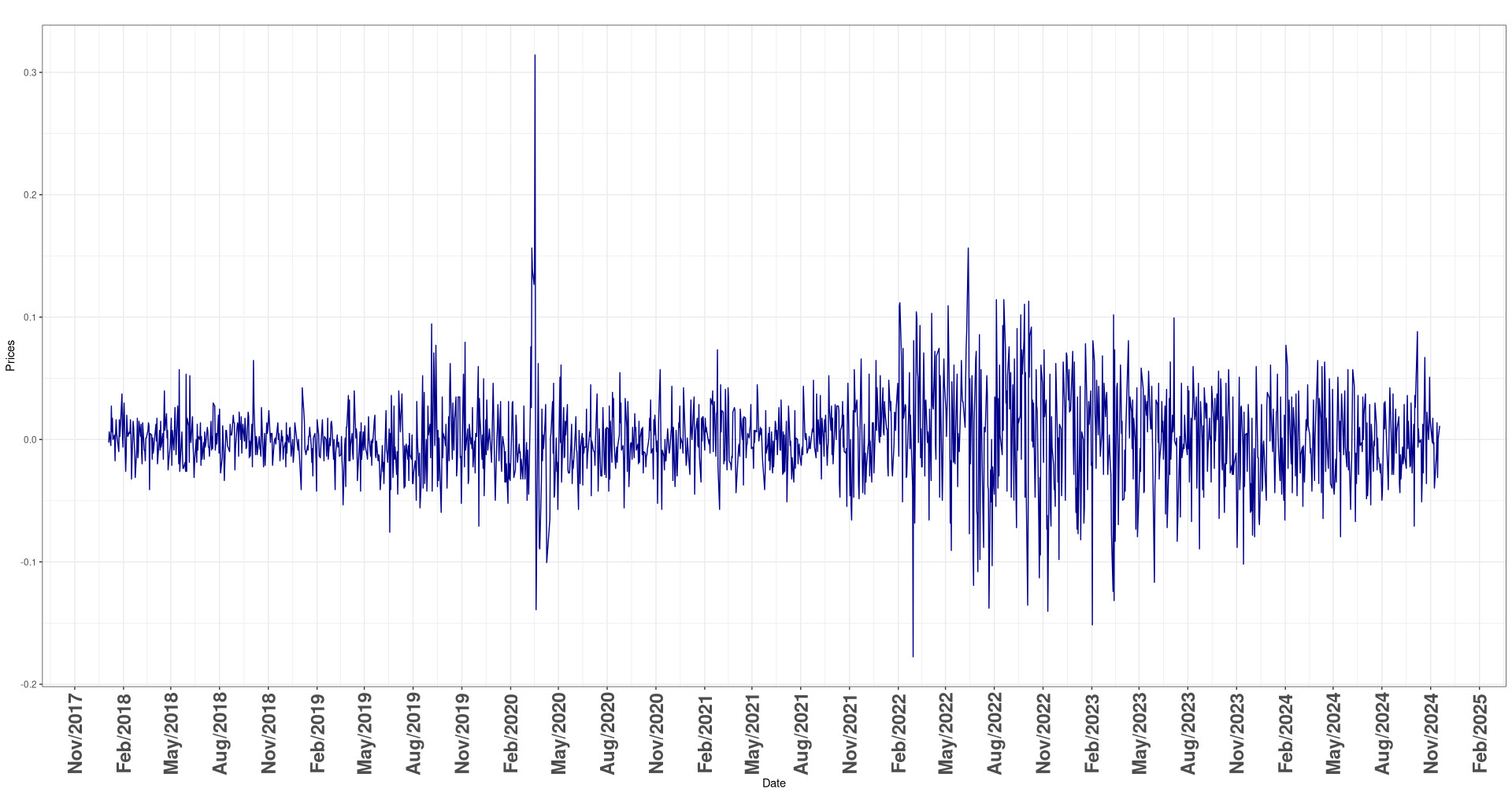}
        \caption{Amundi Responsible Investing - Impact Green Bonds DP }
        \label{fig:fig5}
    \end{subfigure}
    \hfill
    \begin{subfigure}{0.49\textwidth}
        \centering
        \includegraphics[width=\linewidth]{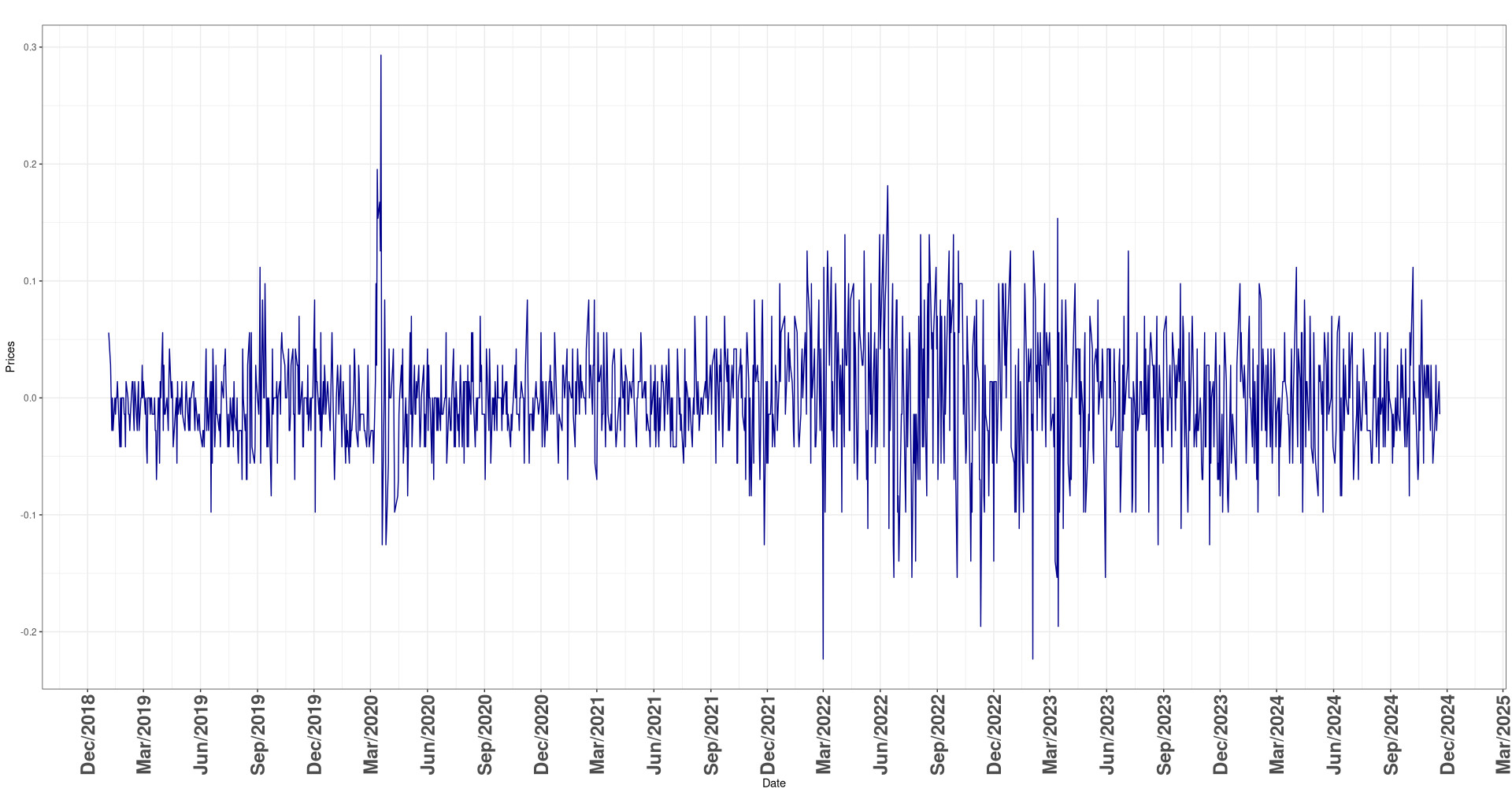}
        \caption{iShares Green Bond Index Fund (IE) D Acc USD Hedged}
        \label{fig:fig6}
    \end{subfigure}
    
    \begin{subfigure}{0.49\textwidth}
        \centering
        \includegraphics[width=\linewidth]{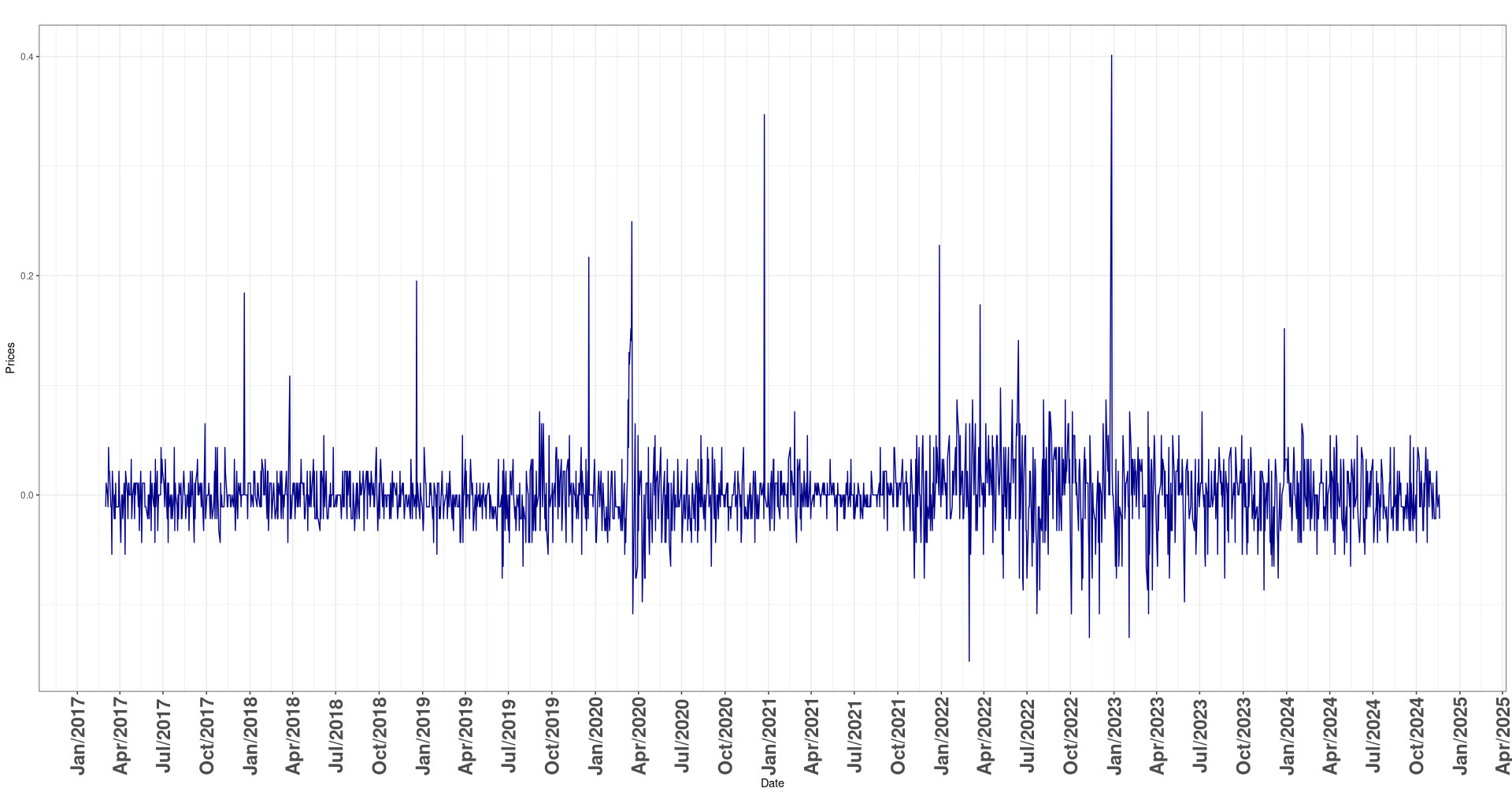}
        \caption{Mirova Global Green Bond Fund Class Y }
        \label{fig:fig7}
    \end{subfigure}
    \hfill
    \begin{subfigure}{0.49\textwidth}
        \centering
        \includegraphics[width=\linewidth]{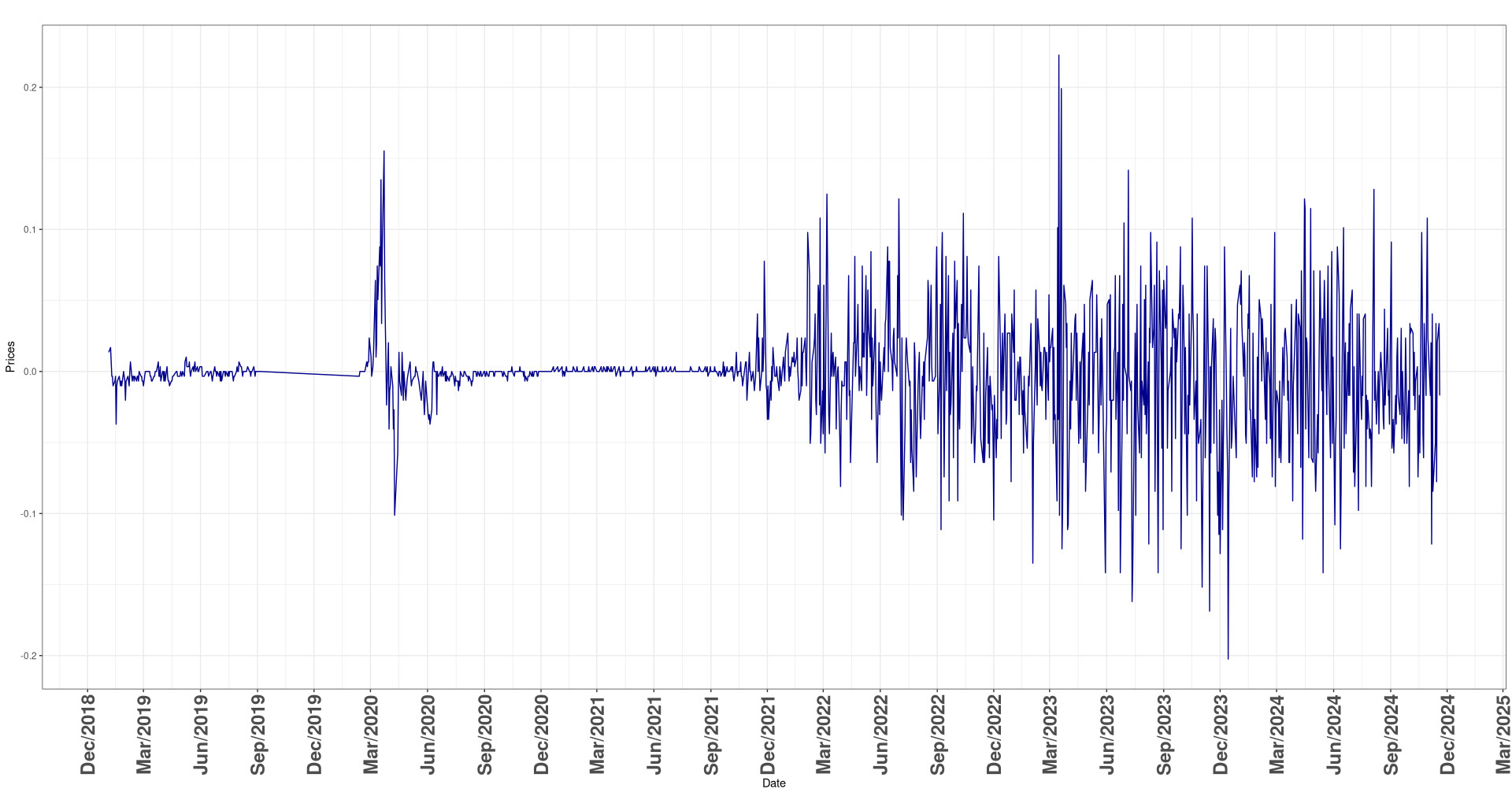}
        \caption{Swiss Life Funds (F) Bond Floating Rates F}
        \label{fig:fig8}
    \end{subfigure}
    
    \caption{Plots of the four Green Bond Funds price returns.}
    \label{fig:return}
\end{figure}

For each Green Bond fund, we calculate the logarithmic returns using the following mathematical equation: $ R_t = \log (z_{(t)}) - \log(z_{(t-1)})\label{eq:1}$. The return series can be viewed in Figure \ref{fig:return} for the four funds mentioned above. The MFDFA method was implemented to close for all 35 Green Bonds price changes, where local trends are fitted with a second-degree polynomial $m = 2$. Green Bonds funds with satisfactorily long times $> ~ 700$ days are chosen since they can last for closing anywhere between a few months and years. 



{\small
\begin{longtable}{cccc}
 \caption{\textbf{List of the 35 Green Bonds Funds analyzed in this paper.}}
 \label{tab:tab1}\\
\hline
No.& Fund   & Code & Currency\\
   \hline
\endfirsthead
\hline
No.& Fund   & Code & Currency\\
\hline
\endhead
\hline
\multicolumn{4}{r}{Continued on next page} \\
\endfoot
\hline
\endlastfoot
  1  & Amundi Crédit Green Bonds IC & FR0010001214 & EUR \\
  2  & Amundi Responsible Investing - Impact Green Bonds DP C & FR0013188745 & EUR \\
  3  & Amundi Responsible Investing - Impact Green Bonds I2 C & FR0013188737 & EUR \\
  4  & Amundi Responsible Investing - Impact Green Bonds R C & FR0013332160 & EUR \\
  5  & Amundi Responsible Investing - Impact Green Bonds R1 C & FR0013275245 & EUR \\
  6  & Amundi Responsible Investing - Impact Green Bonds R2 D & FR0013275252 & EUR \\
  7  & BRED Green Bonds D & FR0010904854 & EUR \\
  8  & Calvert Green Bond Fund Class A & CGAFX & USD \\
  9  & Calvert Green Bond Fund Class I & CGBIX & USD \\
 10  & Calvert Green Bond Fund Class R6 & CBGRX & USD \\
 11  & CM-AM SICAV - CM-AM Green Bonds IC & FR0013246550 & EUR \\
 12  & Federal Global Green Bonds I & FR0010207548 & EUR \\
 13  & Federal Global Green Bonds P & FR0007394846 & EUR \\
 14  & Green and Impact Bond France B & FR0013387107 & EUR \\
 15  & Green Bonds Investments P & FR0010734442 & EUR \\
 16  & iShares Green Bond Index Fund (IE) Class D Hedged Acc EUR & IE00BD0DT578 & EUR \\
 17  & iShares Green Bond Index Fund (IE) D Acc CHF Hedged & IE00BD8QG026 & CHF \\
 18  & iShares Green Bond Index Fund (IE) D Acc GBP Hedged & IE00BD5GZQ41 & GBP \\
 19  & iShares Green Bond Index Fund (IE) D Acc USD Hedged & IE00BD8QG463 & USD \\
 20  & iShares Green Bond Index Fund (IE) D Dist CHF Hedged & IE00BD8QG133 & CHF \\
 21  & iShares Green Bond Index Fund (IE) Flexible Acc EUR & IE00BD0DT685 & EUR \\
 22  & iShares Green Bond Index Fund (IE) Flexible Dist EUR & IE00BDFSQF67 & EUR \\
 23  & iShares Green Bond Index Fund (IE) Institutional Acc EUR & IE00BD0DT792 & EUR \\
 24  & Ircantec Green Bonds Amundi AM C/D & FR0007014006 & EUR \\
 25  & Mirova Global Green Bond Fund Class A & MGGAX & USD \\
 26  & Mirova Global Green Bond Fund Class N & MGGNX & USD \\
 27  & Mirova Global Green Bond Fund Class Y & MGGYX & USD \\
 28  & Nuveen Green Bond Fund A Class & TGROX & USD \\
 29  & Nuveen Green Bond Fund I Class & TGRKX & USD \\
 30  & Nuveen Green Bond Fund Premier Class & TGRMX & USD \\
 31  & Nuveen Green Bond Fund Premier Class & TGRLX & USD \\
 32  & Nuveen Green Bond Fund R6 Class & TGRNX & USD \\
 33  & Ofi Invest Green Bonds Euro VYV OBLIGATION EURO & FR0013421567 & EUR \\
 34  & Swiss Life Funds (F) Bond Floating Rates F & FR0013333127 & EUR \\
 35  & Swiss Life Funds (F) Green Bonds Impact I & FR0013123999 & EUR \\
    
\end{longtable}
}

\section{RESULTS AND DISCUSSION}\label{sec:result}

In this section, we measured the complexity of the singularity spectra to understand the multifractal behavior of the thirty-five Green Bonds funds shown in Table \ref{tab:tab1} in their original form and their shuffled form.
The randomization was performed by the mathematical formula $r = 10.000$ x $x_i$, being repeated by $1000$ different random seeds as it was applied by Stosic \textit{et al.} (2020) \cite{stosic2020multifractal} and
Nascimento \textit{et al.} (2022) \cite{nascimento2022covid}. From the original and randomized returns, it is possible to observe the causes of multifractality present in the long-range correlation processes and or probability density function \cite{dos2021mixture}.

Figure \ref{fig:st} shows four Green Bonds funds times series, that figure \ref{fig:fig1} and \ref{fig:fig2} have similar behaviour rising between September 2018 and February 2022, and a trend of fall after that date. Figure \ref{fig:fig3} difference of figures \ref{fig:fig1} and \ref{fig:fig2} from January 2017 to November 2018, after this period presents a remarkable similarity with the other funds described above, already the bottom that most visually differs from the others is the bottom represented by figure \ref{fig:fig4} that presents a moment of stability between December 2019 to March 2020, introducing one that was recovered in April and maintaining a stability until December 2021, decaying until June 2022 and showing uptrend.

\begin{figure}[H]
    \centering
    \begin{subfigure}{0.495\textwidth}
        \centering
        \includegraphics[width=\linewidth]{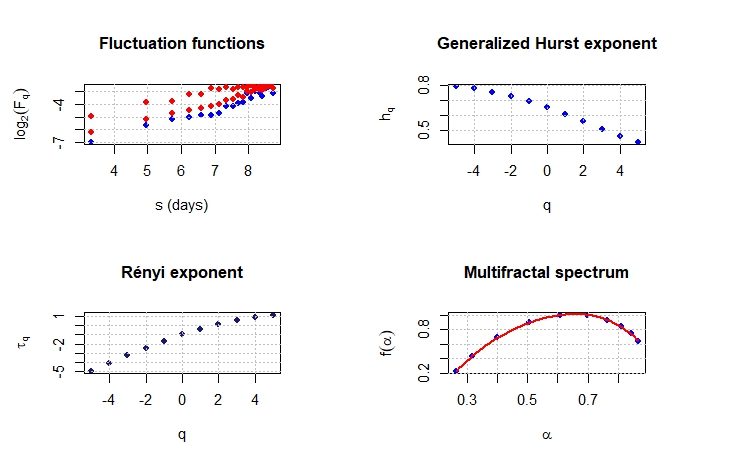}
        \caption{Amundi Responsible Investing - Impact Green Bonds DP }
        \label{fig:mfdfa1}
    \end{subfigure}
    \hfill
    \begin{subfigure}{0.495\textwidth}
        \centering
        \includegraphics[width=\linewidth]{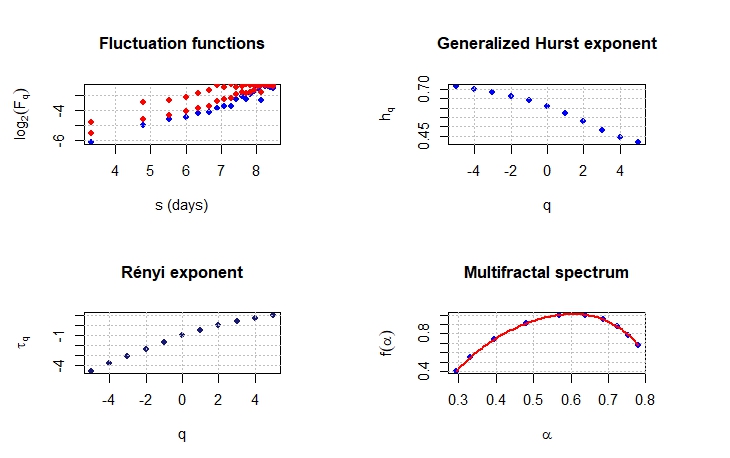}
        \caption{iShares Green Bond Index Fund (IE) D Acc USD Hedged}
        \label{fig:mfdfa2}
    \end{subfigure}
    
    \begin{subfigure}{0.495\textwidth}
        \centering
        \includegraphics[width=\linewidth]{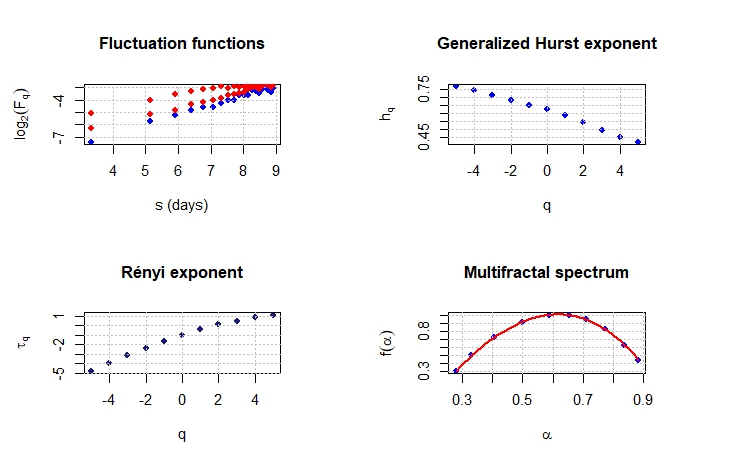}
        \caption{Mirova Global Green Bond Fund Class Y }
        \label{fig:mfdfa3}
    \end{subfigure}
    \hfill
    \begin{subfigure}{0.495\textwidth}
        \centering
        \includegraphics[width=\linewidth]{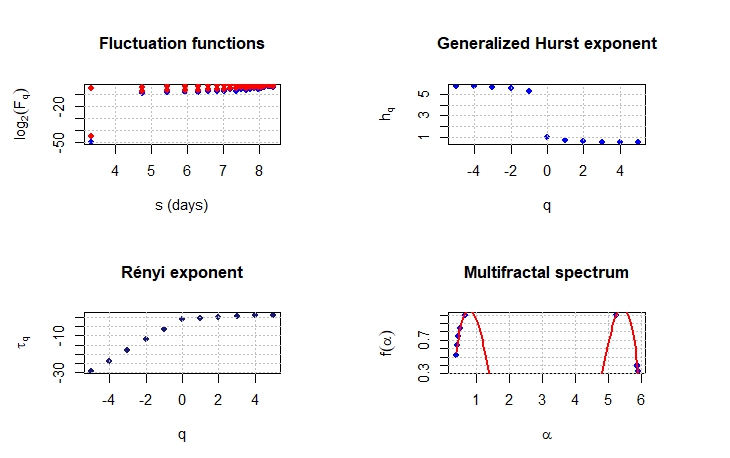}
        \caption{Swiss Life Funds (F) Bond Floating Rates F}
        \label{fig:mfdfa4}
    \end{subfigure}
    
    \caption{Graphs of MFDFA for the time series of returns of four Green Bond Funds.}
    \label{fig:mfdfa}
\end{figure}

Figure \ref{fig:mfdfa} shows the results obtained for four Green Bonds funds, coming from the returns of the original series, where we can see the functions of fluctuation with $q =- 10$, $q=0$, and $q = 10$ demonstrating predictability in the series. The behavior of the exponent of Rényi demonstrates the multifractal behavior of the series. As it can be seen, concave forms, in Figures \ref{fig:mfdfa1}, \ref{fig:mfdfa2}, and \ref{fig:mfdfa3}, corroborate the multifractality of the series and evidence the Green Bond fund FR0013333127 displayed in the Figure \ref{fig:mfdfa4} as an outlier with a strong multifractal trend. 

Due to its behavior represented in Figures \ref{fig:fig4} and\ref{fig:fig8} and the values of its parameters, fund FR0013333127 was an outlier among the 35 Green Bond Funds, presenting an extreme behavior not forming a multifractal spectrum graph displayed in the bell-shaped Figure \ref{fig:mfdfa4} as can be seen in Figures \ref{fig:mfdfa1}, \ref{fig:mfdfa2} and 
\ref{fig:mfdfa3}. 

Table \ref{tab:tab2} shows the parameters obtained using MFDFA in the return price of the series of 35 Green Bond Funds.
The values of Table \ref{tab:tab2} all funds have $\alpha_{0}$ > $0.5$, indicating a persistent behavior, i.e., if an upward or downward trend occurred in the past is likely to occur in the future, we can say that the Green Bonds Funds have long-term memory, that past returns strongly influence future values, not being a random process only, but possessing memory. 

Green Bonds funds mostly have large values of $W$, presenting a high multifractality and pointing to different patterns of variation over time. It does not have a simple linear pattern, but it points to different scales of variation. Being a non-uniform memory, it may vary over time.
Observed that the values of W for the randomized series show a decrease in the width of the multifractal spectrum, having decreased predictability of the series.  
The asymmetry of Green Bond funds is evident in the $r<1$ range, indicating large fluctuations in multifractality and an asymmetric spectrum to the left and demonstrating a large variation between the returns generated by the funds. 

The results of Table \ref{tab:tab2} show that the Green Bonds funds have behavior similar to other financial assets that were analyzed through the MFDFA method, such as cryptocurrencies \cite{stovsic2015multifractal}, \cite{stosic2019multifractal} and agricultural commodities \cite{stosic2020multifractal}, \cite{nascimento2022covid}, stock market indices \cite{zunino2008multifractal}, \cite{stovsic2015multifractal2}, presenting a long-range persistence behavior, marked by very diverse fluctuation and large sudden shocks.
{\small
\begin{longtable}{cccccccccc}
\caption{\textbf{Results of Multifractal Parameters Green Bonds Funds.}}\label{tab:tab2}\\
\hline 
Funds  & $\alpha_0$ & W & r  &$\alpha_0$ (Randomized) & W (Randomized) & r (Randomized) &$\Delta\alpha_0$ &$\Delta$W &$\Delta$r\\
\hline
\endfirsthead

\hline
   Funds  & $\alpha_0$ & W & r  &$\alpha_0$ (Randomized) & W (Randomized) & r (Randomized) &$\Delta\alpha_0$ &$\Delta$W &$\Delta$r\\
      \hline
\endhead
\hline
\multicolumn{10}{r}{Continued on next page} \\
\endfoot
\hline
\endlastfoot 
 FR0010001214 & 0.75 & 0.89 & 0.47 & 0.64 & 0.67 & 0.52 & 0.11 & 0.22 & -0.05 \\
 FR0013188745 & 0.65 & 0.61 & 0.55 & 0.47 & 0.21 & 0.89 & 0.18 & 0.39 & -0.34 \\
 FR0013188737 & 0.65 & 0.61 & 0.63 & 0.51 & 0.22 & 0.38 & 0.14 & 0.39 & 0.24 \\
 FR0013332160  & 0.65 & 0.57 & 0.79 & 0.49 & 0.13 & 0.87 & 0.15 & 0.44 & -0.08 \\ 
 FR0013275245 & 0.65 & 0.58 & 0.51 & 0.47 & 0.21 & 0.82 & 0.18 & 0.38 & -0.31 \\ 
 FR0013275252 & 0.64 & 0.62 & 0.62 & 0.49 & 0.23 & 0.44 & 0.15 & 0.39 & 0.18 \\ 
 FR0010904854 & 0.79 & 1.18 & 1.63 & 0.54 & 0.74 & 0.53 & 0.25 & 0.44 & 1.10 \\ 
 CGAFX & 0.61 & 0.56 & 0.47 & 0.52 & 0.29 & 1.05 & 0.10 & 0.27 & -0.58 \\ 
 CGBIX  & 0.61 & 0.57 & 0.48 & 0.52 & 0.30 & 1.20 & 0.10 & 0.27 & -0.72 \\ 
 CBGRX & 0.64 & 0.46 & 0.47 & 0.54 & 0.35 & 1.93 & 0.10 & 0.11 & -1.47 \\ 
 FR0013246550 & 0.68 & 0.54 & 1.02 & 0.55 & 0.21 & 1.31 & 0.13 & 0.33 & -0.29 \\ 
 FR0010207548 & 0.51 & 0.36 & 0.98 & 0.49 & 0.18 & 0.89 & 0.02 & 0.17 & 0.09 \\ 
 FR0007394846 & 0.51 & 0.37 & 1.02 & 0.49 & 0.19 & 0.94 & 0.02 & 0.18 & 0.08 \\ 
 FR0013387107 & 0.65 & 0.70 & 0.73 & 0.50 & 0.12 & 0.42 & 0.15 & 0.58 & 0.31 \\ 
 FR0010734442  & 0.80 & 0.72 & 0.91 & 0.57 & 0.60 & 0.97 & 0.23 & 0.13 & -0.06 \\
 IE00BD0DT578 & 0.63 & 0.51 & 0.57 & 0.53 & 0.22 & 1.03 & 0.10 & 0.29 & -0.45 \\ 
 IE00BD8QG026 & 0.63 & 0.45 & 0.57 & 0.50 & 0.29 & 1.10 & 0.13 & 0.16 & -0.53 \\ 
 IE00BD5GZQ41 & 0.62 & 0.43 & 0.76 & 0.46 & 0.23 & 2.93 & 0.16 & 0.20 & -2.18 \\ 
 IE00BD8QG463 & 0.60 & 0.49 & 0.57 & 0.57 & 0.19 & 1.14 & 0.04 & 0.30 & -0.57 \\ 
 IE00BD8QG133 & 0.62 & 0.46 & 0.65 & 0.53 & 0.33 & 1.04 & 0.09 & 0.13 & -0.40 \\ 
 IE00BD0DT685  & 0.63 & 0.54 & 0.65 & 0.54 & 0.23 & 0.90 & 0.09 & 0.31 & -0.25 \\ 
 IE00BDFSQF67 & 0.55 & 0.40 & 0.52 & 0.50 & 0.22 & 5.27 & 0.05 & 0.18 & -4.75 \\ 
 IE00BD0DT792 & 0.63 & 0.53 & 0.63 & 0.54 & 0.23 & 1.25 & 0.09 & 0.30 & -0.62 \\ 
 FR0007014006 & 0.70 & 0.85 & 1.09 & 0.54 & 0.24 & 1.27 & 0.16 & 0.60 & -0.18 \\ 
 MGGAX & 0.62 & 0.51 & 0.49 & 0.56 & 0.33 & 0.44 & 0.07 & 0.18 & 0.05 \\
 MGGNX & 0.62 & 0.61 & 0.79 & 0.55 & 0.34 & 0.56 & 0.07 & 0.27 & 0.22 \\ 
 MGGYX & 0.62 & 0.60 & 0.77 & 0.55 & 0.31 & 0.36 & 0.08 & 0.29 & 0.41 \\
 TGROX & 0.65 & 0.75 & 0.36 & 0.54 & 0.28 & 0.54 & 0.10 & 0.47 & -0.18 \\ 
 TGRKX & 0.65 & 0.78 & 0.42 & 0.55 & 0.25 & 0.54 & 0.10 & 0.53 & -0.12 \\ 
 TGRMX & 0.65 & 0.79 & 0.43 & 0.54 & 0.26 & 0.57 & 0.11 & 0.52 & -0.14 \\ 
 TGRLX & 0.65 & 0.78 & 0.42 & 0.56 & 0.26 & 0.56 & 0.09 & 0.51 & -0.14 \\ 
 TGRNX & 0.65 & 0.78 & 0.42 & 0.55 & 0.27 & 0.54 & 0.10 & 0.51 & -0.12 \\
 FR0013421567 & 0.53 & 0.31 & 0.56 & 0.60 & 0.18 & 1.38 & -0.07 & 0.12 & -0.82 \\
 FR0013333127 & 2.95 & 5.52 & 1.16 & 0.58 & 0.31 & 7.71 & 2.37 & 5.21 & -6.55 \\ 
 FR0013123999  & 0.98 & 1.18 & 0.84 & 0.54 & 0.40 & 4.98 & 0.45 & 0.79 & -4.14 \\ 
\end{longtable}
}
\section{CONCLUSION}\label{sec:concluse}
This paper applies the MFDFA method and analyzes its multifractal properties to investigate its behavior and dynamics in thirty-five Green Bond Funds. We use the time series of logarithmic returns from closing prices.  Similar parameters were found to other financial assets, such as cryptocurrencies and commodities. The series showed general persistence behavior of $\alpha_{0} > 0.5$, indicating that the series has predictability, marked by greater complexity, of heterogeneous behavior, causing great shocks.
The FR0013333127 fund had a different behavior from other funds, presenting an extreme multifractality with a high degree of predictability.

Our work presents a pioneering multifractal analysis study on the Green Bonds funds, a financial instrument that has been gaining visibility from the climate change faced by the world \cite{lee2024climate}, confirming that they have very similar characteristics to other financial assets, such as cryptocurrencies and commodities. Contributing to a better collection of funds for the application in Green Bonds in the financial market.

Through this article, asset managers can understand that investing in Green Bonds involves acquiring securities that have the characteristics of other assets and encouraging Green Bonds in their strategies to increase their profits, mitigate losses, and contribute to the development of sustainable ventures that aim to reduce climate change \cite{lee2024climate} has caused disturbances around the planet.

Future studies should be directed to the investigation of multifractal dynamics and correlations of price and volume changes with the use of sliding windows. We believe that this study contributed to the understanding of the behavior of Green Bond funds in the financial market. 

\section*{ACKNOWLEDGMENTS}
Coordination for the Improvement of Higher Education Personnel — Brazil (CAPES) — Finance Code 001 by financial support for the development of this research.

\bibliography{bibliografia2}

@article{figliola2007multifractal,
  title={Multifractal detrented fluctuation analysis of tonic-clonic epileptic seizures},
  author={Figliola, A and Serrano, E and Rosso, OA},
  journal={The European physical journal special topics},
  volume={143},
  number={1},
  pages={117--123},
  year={2007},
  publisher={Springer}
}

@article{kantelhardt2006long,
  title={Long-term persistence and multifractality of precipitation and river runoff records},
  author={Kantelhardt, Jan W and Koscielny-Bunde, Eva and Rybski, Diego and Braun, Peter and Bunde, Armin and Havlin, Shlomo},
  journal={Journal of Geophysical Research: Atmospheres},
  volume={111},
  number={D1},
  year={2006},
  publisher={Wiley Online Library}
}

@article{telesca2006measuring,
  title={Measuring multifractality in seismic sequences},
  author={Telesca, Luciano and Lapenna, Vincenzo},
  journal={Tectonophysics},
  volume={423},
  number={1-4},
  pages={115--123},
  year={2006},
  publisher={Elsevier}
}

@article{matia2003multifractal,
  title={Multifractal properties of price fluctuations of stocks and commodities},
  author={Matia, Kaushik and Ashkenazy, Yosef and Stanley, H Eugene},
  journal={Europhysics letters},
  volume={61},
  number={3},
  pages={422},
  year={2003},
  publisher={IOP Publishing}
}

@article{zunino2008multifractal,
  title={A multifractal approach for stock market inefficiency},
  author={Zunino, Luciano and Tabak, Benjamin Miranda and Figliola, Alejandra and P{\'e}rez, Dar{\'\i}o G and Garavaglia, Mario and Rosso, Osvaldo A},
  journal={Physica A: Statistical Mechanics and its Applications},
  volume={387},
  number={26},
  pages={6558--6566},
  year={2008},
  publisher={Elsevier}
}

@article{oh2012multifractal,
  title={A multifractal analysis of Asian foreign exchange markets},
  author={Oh, Gabjin and Eom, Cheoljun and Havlin, Shlomo and Jung, W -S and Wang, Fengzhong and Stanley, H Eugene and Kim, Seunghwan},
  journal={The European Physical Journal B},
  volume={85},
  pages={1--6},
  year={2012},
  publisher={Springer}
}

@article{stovsic2015multifractal,
  title={Multifractal analysis of managed and independent float exchange rates},
  author={Stosic, Darko and Stosic, Dusan and Stosic, Tatijana and Stanley, H Eugene},
  journal={Physica A: Statistical Mechanics and its Applications},
  volume={428},
  pages={13--18},
  year={2015},
  publisher={Elsevier}
}

@article{kantelhardt2002multifractal,
  title={Multifractal detrended fluctuation analysis of nonstationary time series},
  author={Kantelhardt, Jan W and Zschiegner, Stephan A and Koscielny-Bunde, Eva and Havlin, Shlomo and Bunde, Armin and Stanley, H Eugene},
  journal={Physica A: Statistical Mechanics and its Applications},
  volume={316},
  number={1-4},
  pages={87--114},
  year={2002},
  publisher={Elsevier}
}

@article{nascimento2022covid,
  title={COVID-19 Influence over Brazilian Agricultural Commodities and Dollar--Real Exchange},
  author={Nascimento, Kerolly Kedma Felix Do and Santos, F{\'a}bio Sandro Dos and Nascimento, Kellyma Kellyashin Felix Do and Nascimento, Kenikywaynne Kerowaynne Felix Do and J{\'u}nior, Silvio Fernando Alves Xavier and Jale, Jader Silva and Ferreira, Tiago AE},
  journal={Fractals},
  volume={30},
  number={06},
  pages={2250100},
  year={2022},
  publisher={World Scientific}
}

@article{choi2021analysis,
  title={Analysis of stock market efficiency during crisis periods in the US stock market: Differences between the global financial crisis and COVID-19 pandemic},
  author={Choi, Sun-Yong},
  journal={Physica A: Statistical Mechanics and Its Applications},
  volume={574},
  pages={125988},
  year={2021},
  publisher={Elsevier}
}

@article{akinsusi2022nonlinear,
  title={Nonlinear dynamics and multifractal analysis of minimum--maximum temperature and solar radiation over Lagos State, Nigeria},
  author={Akinsusi, Joshua and Ogunjo, Samuel and Fuwape, Ibiyinka},
  journal={Acta Geophysica},
  volume={70},
  number={5},
  pages={2171--2178},
  year={2022},
  publisher={Springer}
}

@article{wang2022impact,
  title={The impact of the COVID-19 pandemic on the energy market--A comparative relationship between oil and coal},
  author={Wang, Qiang and Yang, Xuan and Li, Rongrong},
  journal={Energy Strategy Reviews},
  volume={39},
  pages={100761},
  year={2022},
  publisher={Elsevier}
}

@article{dos2023multifractal,
  title={Multifractal Analysis of Solar Radiation in the Northeastern Region of Brazil},
  author={Dos Santos, F{\'a}bio Sandro and Do Nascimento, Kerolly Kedma Felix and Jale, Jader Silva and Xavier J{\'u}nior, S{\'\i}lvio Fernando Alves and Ferreira, Tiago AE},
  journal={Fractals},
  volume={31},
  number={03},
  pages={2350026},
  year={2023},
  publisher={World Scientific}
}

@article{stosic2019multifractal,
  title={Multifractal behavior of price and volume changes in the cryptocurrency market},
  author={Stosic, Darko and Stosic, Dusan and Ludermir, Teresa B and Stosic, Tatijana},
  journal={Physica A: Statistical Mechanics and Its Applications},
  volume={520},
  pages={54--61},
  year={2019},
  publisher={Elsevier}
}

@article{stosic2020multifractal,
  title={Multifractal analysis of Brazilian agricultural market},
  author={Stosic, Tatijana and Nejad, Salman Abarghouei and Stosic, Borko},
  journal={Fractals},
  volume={28},
  number={05},
  pages={2050076},
  year={2020},
  publisher={World Scientific}
}

@article{dos2021mixture,
  title={Mixture distribution and multifractal analysis applied to wind speed in the Brazilian Northeast region},
  author={dos Santos, Fabio Sandro and do Nascimento, Kerolly Kedma Felix and da Silva Jale, Jader and Stosic, Tatijana and Marinho, Manoel HN and Ferreira, Tiago AE},
  journal={Chaos, Solitons \& Fractals},
  volume={144},
  pages={110651},
  year={2021},
  publisher={Elsevier}
}

@article{stovsic2015multifractal2,
  title={Multifractal properties of price change and volume change of stock market indices},
  author={Stosic, Dusan and Stosic, Darko and Stosic, Tatijana and Stanley, H Eugene},
  journal={Physica A: Statistical Mechanics and its Applications},
  volume={428},
  pages={46--51},
  year={2015},
  publisher={Elsevier}
}

@article{li2020institutional,
  title={The institutional characteristics of multifractal spectrum of China’s stock market},
  author={Li, Yong and Vilela, Andr{\'e} LM and Stanley, H Eugene},
  journal={Physica A: Statistical Mechanics and its Applications},
  volume={550},
  pages={124129},
  year={2020},
  publisher={Elsevier}
}

@article{lee2024climate,
  title={Climate change 2023 synthesis report summary for policymakers},
  author={Lee, Hoesung and Calvin, Katherine and Dasgupta, Dipak and Krinner, Gerhard and Mukherji, Aditi and Thorne, Peter and Trisos, Christopher and Romero, Jos{\'e} and Aldunce, Paulina and Ruane, Alexander C},
  journal={CLIMATE CHANGE 2023 Synthesis Report: Summary for Policymakers},
  year={2024},
  publisher={Intergovernmental Panel on Climate Change}
}

@report{Michetti2023,
  author    = {Michetti, C. and others},
  title     = {Sustainable Debt Global State of the Market 2022},
  year      = {2023},
  institution = {Climate Bonds Initiative},
}

@article{ebeling2022european,
  title={European investment bank loan appraisal, the EU climate bank?},
  author={Ebeling, Antoine},
  journal={International Economics},
  volume={172},
  pages={203--216},
  year={2022},
  publisher={Elsevier}
}

@article{jiang2011multifractal,
  title={Multifractal detrending moving-average cross-correlation analysis},
  author={Jiang, Zhi-Qiang and Zhou, Wei-Xing},
  journal={Physical Review E—Statistical, Nonlinear, and Soft Matter Physics},
  volume={84},
  number={1},
  pages={016106},
  year={2011},
  publisher={APS}
}

@article{liu1999statistical,
  title={Statistical properties of the volatility of price fluctuations},
  author={Liu, Yanhui and Gopikrishnan, Parameswaran and Stanley, H Eugene and others},
  journal={Physical review e},
  volume={60},
  number={2},
  pages={1390},
  year={1999},
  publisher={APS}
}

@article{lahmiri2018chaos,
  title={Chaos, randomness and multi-fractality in Bitcoin market},
  author={Lahmiri, Salim and Bekiros, Stelios},
  journal={Chaos, solitons \& fractals},
  volume={106},
  pages={28--34},
  year={2018},
  publisher={Elsevier}
}

@article{moudud2024stock,
  title={Stock market efficiency of the BRICS countries pre-, during, and post covid-19 pandemic: a multifractal detrended fluctuation analysis},
  author={Moudud-Ul-Huq, Syed and Rahman, Md Shahriar},
  journal={Computational Economics},
  pages={1--63},
  year={2024},
  publisher={Springer}
}

@article{jian2023green,
  title={Green bonds and green environment: exploring innovative financing mechanisms for environmental project sustainability},
  author={Jian, Yirong},
  journal={Environmental Science and Pollution Research},
  volume={30},
  number={58},
  pages={122293--122303},
  year={2023},
  publisher={Springer}
}

@article{kartal2024dynamic,
  title={Dynamic relationship between green bonds, energy prices, geopolitical risk, and disaggregated level CO2 emissions: evidence from the globe by novel WLMC approach},
  author={Kartal, Mustafa Tevfik and Ta{\c{s}}k{\i}n, Dilvin and K{\i}l{\i}{\c{c}} Depren, Serpil},
  journal={Air Quality, Atmosphere \& Health},
  volume={17},
  number={8},
  pages={1763--1775},
  year={2024},
  publisher={Springer}
}

@article{Kartal2025,
  author    = {Kartal, Mustafa Tevfik and Pata, Ugur Kilic and Alola, Andrew Amaechi},
  title     = {Impact of green bonds on $CO_{2}$ emissions and disaggregated level renewable electricity in China and the United States of America},
  journal   = {Humanities and Social Sciences Communications},
  volume    = {12},
  pages     = {350},
  year      = {2025},
  doi       = {10.1057/s41599-025-04696-0},
  url       = {https://doi-org.ez19.periodicos.capes.gov.br/10.1057/s41599-025-04696-0}
}

@article{stosic2019multifractalmarket,
  title={Multifractal characterization of Brazilian market sectors},
  author={Stosic, Dusan and Stosic, Darko and de Mattos Neto, Paulo SG and Stosic, Tatijana},
  journal={Physica A: Statistical Mechanics and its Applications},
  volume={525},
  pages={956--964},
  year={2019},
  publisher={Elsevier}
}

@article{nejad2021multifractal,
  title={Multifractal analysis of the gold market},
  author={Nejad, Salman Abarghouei and Stosic, Tatijana and Stosic, Borko},
  journal={Fractals},
  volume={29},
  number={01},
  pages={2150010},
  year={2021},
  publisher={World Scientific}
}

@misc{calvet2022finance,
  title={The finance of climate change},
  author={Calvet, Laurent and Gianfrate, Gianfranco and Uppal, Raman},
  journal={Journal of Corporate Finance},
  volume={73},
  pages={102162},
  year={2022},
  publisher={Elsevier}
}

@article{azizi2023noninteger,
  title={Noninteger dimension of seasonal land surface temperature (LST)},
  author={Azizi, Sepideh and Azizi, Tahmineh},
  journal={Axioms},
  volume={12},
  number={6},
  pages={607},
  year={2023},
  publisher={MDPI}
}

@article{de2025multifractal,
  title={Multifractal analysis of braided channel networks using structure functions and fixed-mass measures},
  author={De Michele, Carlo and Primavera, Leonardo and Tomasicchio, Giuseppe R and Lauria, Agostino and Francone, Antonio and Leone, Elisa and Salvadori, Gianfausto and De Bartolo, Samuele},
  journal={Physica A: Statistical Mechanics and its Applications},
  pages={130831},
  year={2025},
  publisher={Elsevier}
}

@article{liu2025analysis,
  title={Analysis of the Multifractal Characteristics of the Chinese Stock Market Based on Deep Wavelet Transform},
  author={Liu, Yafei and Wang, Tonghao and Cattani, Piercarlo and Mei, Shuli},
  journal={Fluctuation and Noise Letters},
  volume={24},
  number={4},
  pages={2550041--51},
  year={2025}
}

@article{rahmani2025unveiling,
  title={Unveiling climate complexity: a multifractal approach to drought, temperature, and precipitation analysis},
  author={Rahmani, Farhang},
  journal={Acta Geophysica},
  volume={73},
  number={3},
  pages={3007--3024},
  year={2025},
  publisher={Springer}
}
\end{document}